\documentclass[journal]{IEEEtran}
\usepackage{amsmath,amssymb,bm,graphicx,tabularx,array,hhline}
\usepackage{cite}
\usepackage{mathabx,mathtools}
\usepackage{mathptmx}
\usepackage[font={footnotesize}]{caption}
\usepackage{subcaption}
\usepackage{xcolor}
\usepackage[hidelinks]{hyperref}

\newcommand\degrees[1]{\ensuremath{#1^\circ}}

\begin{document}

\title{Common Offset Gathers for Efficient Multistatic Beamforming in Frequency Domain}

\author{Marko~Jakovljevic, Xiaohong~Wang, Ion~Candel, Anthony~E.~Samir, and Kai~Thomenius
\thanks{Corresponding author: Marko Jakovljevic, email: \url{mjakovljevic@mgh.harvard.edu}}
\thanks{M.~Jakovljevic, X.~Wang, I.~Candel, A.~E.~Samir and K.~Thomenius are with the Department of Radiology, Massachusetts General Hospital, Boston, MA 02114, USA.}
}

\markboth{}{}

\maketitle

\begin{abstract}
Frequency-domain beamformers reconstruct ultrasound images with far fewer operations than delay-and-sum (DAS).
However, for multistatic synthetic aperture data, they require Fourier transforms over the full transmit-receive aperture and large amounts of memory, limiting their use for volumetric imaging.
Here, we adapt common offset gathers (COG) from seismic imaging to frequency-domain beamforming:
transmit-receive pairs with the same offset are beamformed independently using a bistatic formulation, holding only one offset gather in memory at a time.
We combine COG with the range-Doppler algorithm (RDA) and evaluate COG RDA against DAS and conventional multistatic RDA using simulated, phantom, and \emph{in vivo} acquisitions.
For linear arrays, COG RDA reduced memory requirements by two orders of magnitude relative to conventional RDA and beamformed up to $25\times$ faster than DAS, with comparable image quality.
For a $32\times32$-element matrix array, COG RDA requires only 2-D instead of 4-D Fourier transforms, and reconstructed a volume on a GPU $32\times$ faster and with $67\times$ less energy than DAS.
COG RDA makes frequency-domain multistatic beamforming practical for matrix arrays and power-constrained devices.
\end{abstract}


\section{Introduction}
\label{sec:introduction}
Medical ultrasound (US) increasingly requires beamformers that process large datasets in real time on limited hardware.
Point-of-care ultrasound (POCUS) and handheld systems \cite{breithardt2015hand,haji2021hand} are constrained by size, weight, and power, which limits processing hardware and, in turn, frame rate and image resolution \cite{le2022comparison,chamsi2017handheld,falkowski2020hand}.
Wearable US patches impose even stricter constraints, with limited battery capacity and no active cooling \cite{hu2023wearable}.

High-data-throughput applications challenge even full-size clinical scanners: functional ultrasound imaging requires Doppler ensembles hundreds of frames long \cite{mace2013functional,jakovljevic2020blood},
and fully sampled matrix arrays generate large multi-dimensional channel datasets that must be beamformed in real time \cite{Smith92,vonRamm91b}.
Piezoelectric (PMUT) and capacitive (CMUT) micromachined ultrasound transducers \cite{moisello2024pmut,qiu2015piezoelectric} now enable 2-D apertures with thousands of elements fabricated using scalable semiconductor processes,
but the reconstruction workloads they impose remain difficult to support, especially on portable, battery-powered platforms.

Ultrasound systems typically reconstruct images using the delay-and-sum (DAS) beamformer, which focuses each image pixel individually through interpolation of channel signals at pre-calculated pulse-echo travel times \cite{Cobbold07}.
For synthetic aperture (SA) processing of multistatic datasets, the number of DAS operations scales with the product of the numbers of transmit events, receive elements, and reconstructed pixels \cite{jakovljevic2022adaptation},
so volumetric imaging with fully sampled 2-D arrays can raise the computational load by 4--6 orders of magnitude.
Real-time volumetric DAS therefore relies on hardware acceleration using field-programmable gate arrays (FPGAs) \cite{meacci2016high} or graphics processing units (GPUs) \cite{hyun2019open,yiu2011gpu},
which increases power consumption, heat generation, system cost, and complexity, and limits portability.

Beamforming in the spatial frequency domain offers an alternative to DAS that reduces the number of arithmetic operations, since a single wavefront migration in Fourier-domain can focus signals at many image pixels simultaneously,
enabling block-based beamforming \cite{Cumming2005digital}.
Frequency-domain algorithms originally developed for synthetic aperture radar (SAR) and seismic imaging have been adapted to ultrasound,
including the F-$k$ (Stolt) migration \cite{Garcia13,Moghimirad16,Zhang16}, the range-Doppler algorithm (RDA) \cite{Ylitalo1994,jakovljevic2022adaptation}, and the chirp scaling algorithm (CSA) \cite{zhuang2026fourier}.
These methods can reduce beamforming time by one to two orders of magnitude compared to DAS while preserving image quality \cite{jakovljevic2022adaptation,zhuang2026fourier}.
Extensions to 2-D apertures have been demonstrated for plane-wave \cite{merabet20192} and row-column-addressed (RCA) array \cite{lin2023fourier,guan20253d} acquisitions,
but not for multistatic SA sequences or virtual source (VS) reconstruction from focused transmits and diverging waves,
which involve substantially more transmits and are necessary to achieve high image resolution and uniformity.

One of the key obstacles to widespread adoption of frequency-domain beamformers for multistatic and VS acquisitions is high memory consumption.
Standard implementations \cite{Hunter08,Zhang16,jakovljevic2022adaptation,jakovljevic2026power} require multi-dimensional Fourier transforms (FTs) along both the transmit and receive aperture dimensions of the channel data matrix,
with zero-padding in each dimension (typically on the order of the data size) to avoid interpolation errors during migration \cite{jakovljevic2022adaptation}.
As a result, they can require several gigabytes of memory to reconstruct a single B-mode frame from a 1-D array, and even more for matrix arrays, where additional FFT dimensions make them prohibitive for resource-constrained platforms.

In this work, we adapt the concept of the \emph{common offset gather} (COG), originally developed in geophysical seismic imaging \cite{cary1999common,deregowski2008common}, to reduce the memory requirements of frequency-domain multistatic beamforming, while also reducing beamforming time in practice.
COG groups channel signals from transmit-receive element pairs sharing the same aperture separation (offset) and beamforms each group independently using a bistatic formulation that requires only one-dimensional spatial FFTs.
Because the multistatic channel data matrix is processed one diagonal at a time rather than as a whole, the peak memory required at any point during reconstruction scales with the length of a single diagonal rather than the full matrix size.
The COG concept is applicable to any frequency-domain beamforming algorithm and to a broad range of acquisition sequences, including multistatic and virtual source acquisitions.

Without loss of generality, we demonstrate COG beamforming using RDA for multistatic acquisitions on both 1-D linear arrays and fully sampled 2-D matrix arrays,
and show that COG RDA significantly reduces beamforming time and memory relative to standard multistatic RDA, while maintaining comparable image quality.
For volumetric imaging with 2-D arrays, where standard multistatic RDA would require a 4-D FFT, COG RDA requires only 2-D FFTs, making frequency-domain multistatic processing practically feasible,
with a speedup over DAS that grows with array size \cite{jakovljevic2022adaptation}.

The paper is organized as follows.
Section~\ref{sec:theory} develops the signal models and frequency-domain expressions underlying COG beamforming for 1-D and 2-D apertures.
Section~\ref{sec:methods} describes the simulation setups, phantom acquisitions, image reconstruction pipeline, and performance metrics used to evaluate COG RDA.
Imaging results and efficiency measurements for 2-D B-mode and volumetric imaging are presented in Section~\ref{sec:results}.
The significance of the results and directions for future work are discussed in Section~\ref{sec:discussion}, and concluding remarks are given in Section~\ref{sec:conclusion}.

\section{Theory}
\label{sec:theory}
\subsection{Common Offset Gathers on 1-D Apertures}
\label{subsec:cog_1d}

\begin{figure}[!t]
\centering
\includegraphics[width=\columnwidth,trim={0.5cm 0cm 11.5cm 0cm},clip]{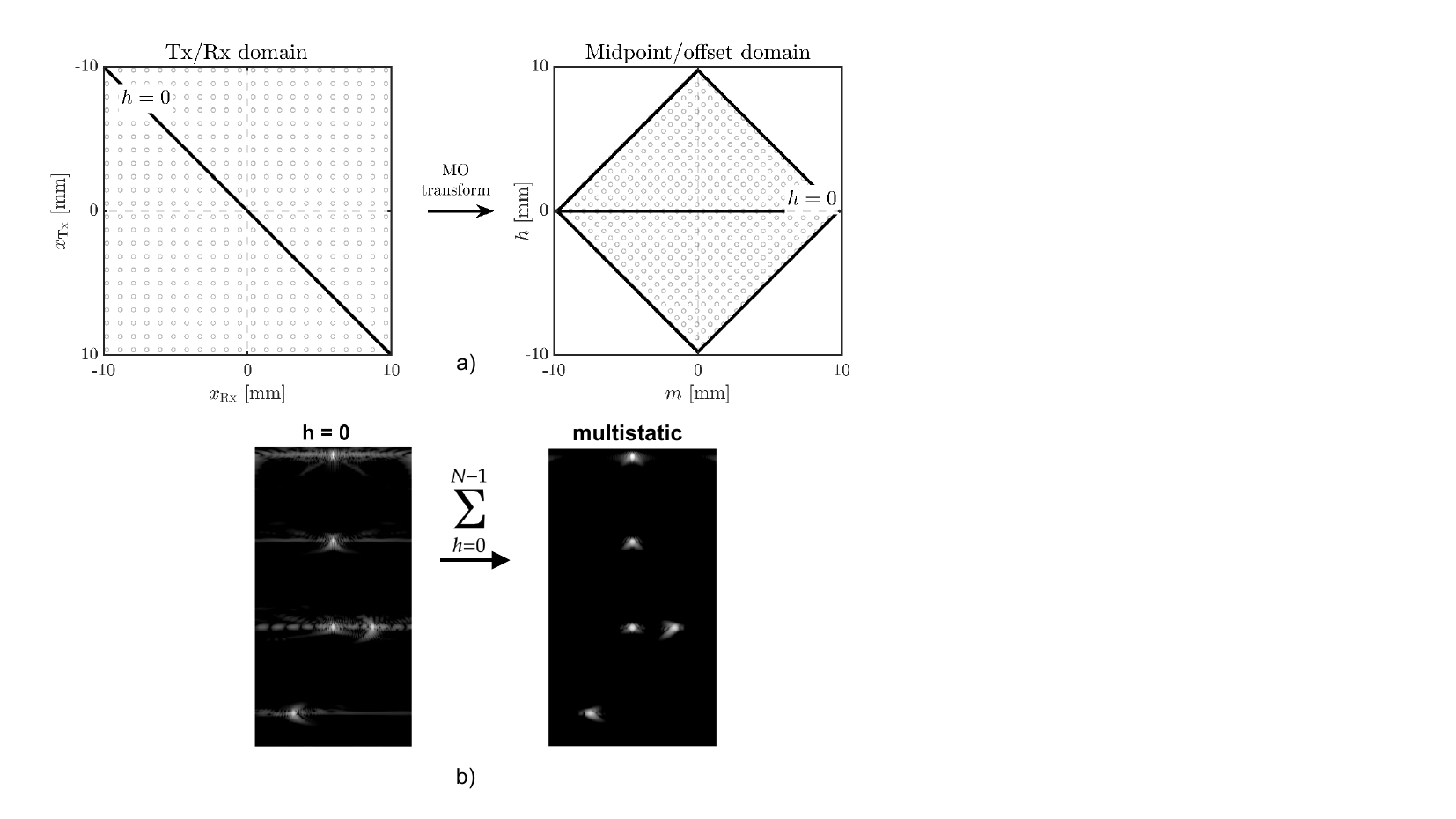}
\caption{Common offset gathers (COG) illustrated for simulated point-target data from a 1-D array.
(a) The multistatic data matrix in Tx/Rx coordinates is reindexed to midpoint-offset (MO) coordinates via the MO transform in Eq.~\eqref{eq:mo_coords_1d}.
The zero-offset diagonal ($h = 0$), corresponding to monostatic acquisition, is highlighted in both domains.
(b) Each COG is beamformed independently using a bistatic signal model, as illustrated for $h = 0$.
Coherent summation of partial images across all $N$ offsets yields the full multistatic image.}
\label{fig:cog_concept}
\end{figure}

Consider a full synthetic aperture (SA) acquisition with a 1-D array of $N$ elements.
The recorded signals form an $N \times N$ data matrix $\mathbf{E}(t)$, where entry $e_{ij}(t)$ is a time series recorded by element $j$ in response to a transmission from element $i$:
\begin{equation}
\mathbf{E}(t) =
\begin{bmatrix}
e_{11}(t) & e_{12}(t) & \cdots & e_{1N}(t) \\
e_{21}(t) & e_{22}(t) &        & \vdots    \\
\vdots    &           & \ddots &           \\
e_{N1}(t) &           &        & e_{NN}(t)
\end{bmatrix}.
\label{eq:data_matrix}
\end{equation}
Common offset gathers (COG) \cite{cary1999common,deregowski2008common} group transmit-receive (Tx/Rx) pairs sharing the same aperture separation, or \emph{offset}.
These pairs lie along the diagonals of $\mathbf{E}(t)$.
Each diagonal is beamformed independently using a bistatic signal model; coherent summation across diagonals yields the full multistatic image.
This is illustrated in Fig.~\ref{fig:cog_concept}.

We introduce midpoint-offset (MO) coordinates to parameterize each diagonal:
\begin{equation}
m = \frac{x_{tx} + x_{rx}}{2}, \qquad h = \frac{x_{tx} - x_{rx}}{2},
\label{eq:mo_coords_1d}
\end{equation}
where $x_{tx}$ and $x_{rx}$ are the transmit and receive element positions, respectively, $m$ is their midpoint, and $h$ is the half-offset.
In MO coordinates, each diagonal of $\mathbf{E}(t)$ corresponds to a fixed $h$, with signals along that diagonal parameterized by $m$.
The multistatic image is formed by coherent accumulation over all diagonals:
\begin{equation}
I(x,z) = \sum_{h} I_h(x,z),
\label{eq:cog_sum}
\end{equation}
where $I_h(x,z)$ is the bistatic image beamformed from the diagonal at offset $h$.
For a 1-D array of $N$ elements, there are $N$ diagonals, and processing them sequentially avoids the need to load the entire data matrix into memory at once.

\subsection{COG Beamforming with the Range-Doppler Algorithm}
\label{subsec:cog_rda_1d}

We combine COG with the Range-Doppler Algorithm (RDA) \cite{Cumming2005digital,jakovljevic2022adaptation}.
For a fixed offset $h$, the baseband signal from a point scatterer at depth $R_0$ and lateral position $x = 0$ is
\begin{equation}
s_0(t,m,h) = A_0\,w_r\bigl(t - (R_{tx}+R_{rx})/c\bigr)\, e^{-j2\pi f_0 (R_{tx}+R_{rx})/c},
\label{eq:signal_mo}
\end{equation}
where $A_0$ is the scatterer amplitude, $w_r$ is the range envelope, $f_0$ is the transmit center frequency, $c$ is the speed of sound, and
\begin{equation}
R_{tx} = \sqrt{R_0^2 + (m+h)^2}, \quad R_{rx} = \sqrt{R_0^2 + (m-h)^2}.
\label{eq:travel_path_1d_complete}
\end{equation}
For fixed $h$, a second-order Taylor expansion of the total travel path around $m = 0$ gives
\begin{equation}
R_{tx} + R_{rx} \approx
  2\sqrt{R_0^2+h^2} + \frac{R_0^2}{(R_0^2+h^2)^{3/2}}\,m^2 + \mathcal{O}(m^4).
\label{eq:travel_path_1d}
\end{equation}
Eq.~\eqref{eq:travel_path_1d} separates the midpoint and offset components of the travel path, enabling a closed-form analytic expression in the frequency domain.

For a constant $h$, applying a 1-D Fourier transform over $m$ and evaluating the resulting integral via the Principle of Stationary Phase (POSP; see Appendix~\ref{app:posp_1d}) yields the midpoint modulation rate
\begin{equation}
K_m = \frac{2f_0}{c} \cdot \frac{R_0^2}{\bigl(R_0^2 + h^2\bigr)^{3/2}},
\label{eq:Km}
\end{equation}
and the bistatic signal model in the range-Doppler (RD) domain:
\begin{align}
S_{rd}(t,k_m) &= A_0\,w_r\bigl(t - t(k_m)\bigr) \nonumber \\
  &\quad\cdot \exp\!\left(- j\frac{4\pi f_0}{c}\sqrt{R_0^2+h^2}
  + j\pi\frac{k_m^2}{K_m}\right),
\label{eq:signal_rd} \\
t(k_m) &= \frac{2\sqrt{R_0^2+h^2}}{c} + \frac{k_m^2}{2K_m f_0},
\label{eq:travel_time_rd}
\end{align}
where $k_m$ is the midpoint spatial frequency.
The term $t(k_m)$ in \eqref{eq:travel_time_rd} describes the range cell migration (RCM), i.e. the wavefront variation with $k_m$.
Range cell migration correction (RCMC) is implemented by 1-D interpolation, followed by matched filtering and a 1-D IFFT \cite{jakovljevic2022adaptation}.
Because the Fourier transform is applied only along the midpoint dimension $m$, COG RDA requires a single 1-D FFT per diagonal,
compared to the 2-D FFT over the full transmit-receive aperture required by standard multistatic RDA \cite{jakovljevic2022adaptation}.

\begin{figure*}[!t]
\centering
\includegraphics[width=0.95\textwidth]{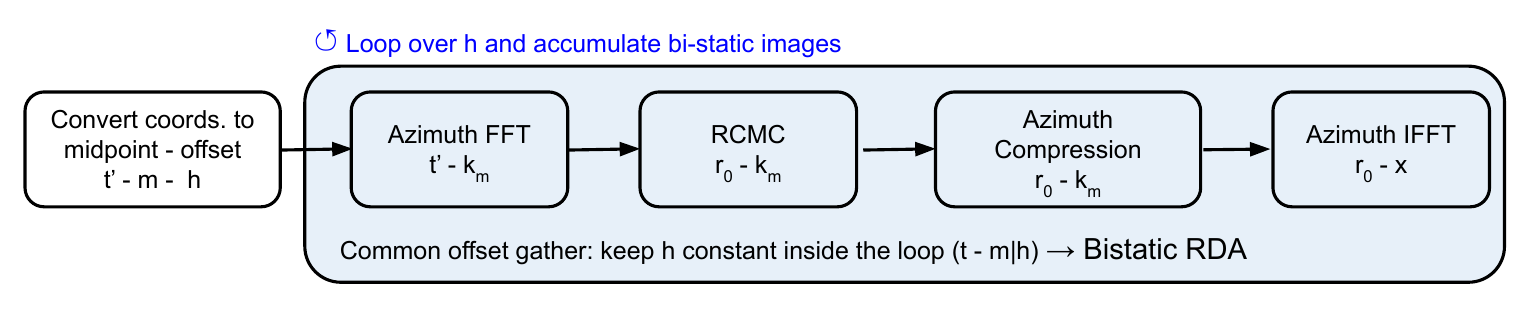}
\caption{Key processing steps of the COG RDA beamformer.
Multistatic channel signals are first reorganized into midpoint/offset (MO) coordinates, grouping element pairs by common offset (corresponding to diagonals of the data matrix illustrated in Fig.~\ref{fig:cog_concept}).
A bistatic version of RDA is applied independently to each offset group.
Partial images from all offset groups are coherently accumulated to produce the final multistatic image.
Labels below each processing block indicate the data dimensions at that stage.}
\label{fig:cog_rda_flowchart}
\end{figure*}

\subsection{COG RDA on 2-D Apertures}
\label{subsec:cog_rda_2d}

\begin{figure}[!t]
\centering
\includegraphics[width=0.75\columnwidth,trim={0 0 0 22bp},clip]{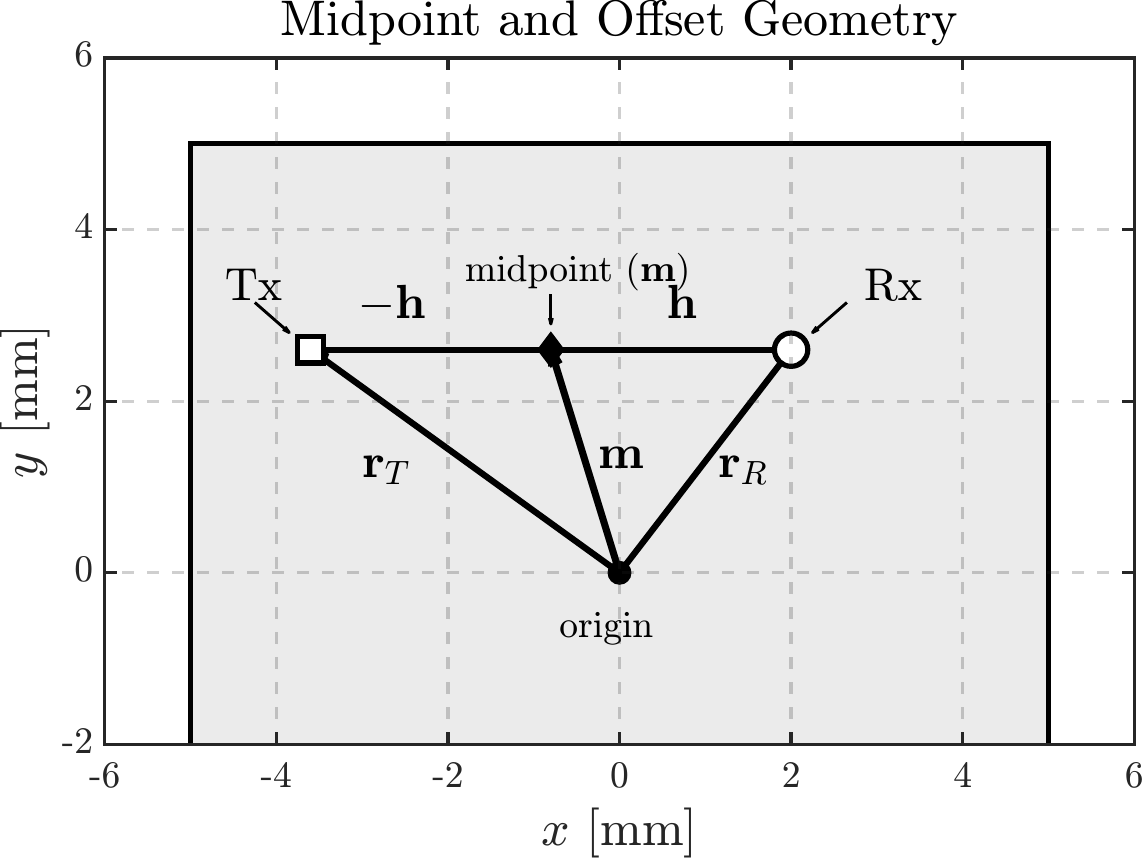}
\caption{Midpoint-offset (MO) coordinates for a 2-D aperture, as defined in \eqref{eq:mo_coords_2d}.
Each Tx/Rx element pair is parameterized by its midpoint $\mathbf{m}$ and half-offset $\mathbf{h}$.}
\label{fig:mo_2d}
\end{figure}

The COG framework extends easily to 2-D apertures by replacing scalar coordinates with 2-D vectors.
For a fully sampled matrix array with transmit and receive element positions $\mathbf{r}_{tx}, \mathbf{r}_{rx} \in \mathbb{R}^2$, the midpoint and offset vectors are
\begin{equation}
\mathbf{m} = \frac{\mathbf{r}_{tx} + \mathbf{r}_{rx}}{2}, \qquad
\mathbf{h} = \frac{\mathbf{r}_{tx} - \mathbf{r}_{rx}}{2},
\label{eq:mo_coords_2d}
\end{equation}
with $\mathbf{m},\mathbf{h} \in \mathbb{R}^2$.
The multistatic dataset from a 2-D aperture forms a 4-D data tensor in $(x_{tx}, y_{tx}, x_{rx}, y_{rx})$ \cite{Savord03}.
In MO coordinates, this tensor is partitioned into constant-$\mathbf{h}$ groups, each beamformed independently over the midpoint vector $\mathbf{m} = (x_m, y_m)$ using a bistatic volumetric RDA.
This geometry is illustrated in Fig.~\ref{fig:mo_2d}.

For fixed $\mathbf{h}$, the travel path to a scatterer at depth $R_0$ admits the second-order approximation (see Appendix~\ref{app:posp_2d})
\begin{equation}
R_{tx}(\mathbf{m},\mathbf{h}) + R_{rx}(\mathbf{m},\mathbf{h})
  \approx 2\sqrt{D} + \mathbf{m}^\top \mathbf{A}\,\mathbf{m},
\label{eq:travel_path_2d}
\end{equation}
where $D := R_0^2 + \|\mathbf{h}\|^2$ and
\begin{equation}
\mathbf{A} := \frac{1}{\sqrt{D}}\,\mathbf{I} - \frac{1}{D^{3/2}}\,\mathbf{h}\mathbf{h}^\top.
\label{eq:A_matrix}
\end{equation}
Applying the 2-D POSP to the Fourier transform of the signal over $\mathbf{m}$ (Appendix~\ref{app:posp_2d}), and using the Sherman-Morrison formula to invert $\mathbf{A}$,
\begin{equation}
\mathbf{A}^{-1} = \sqrt{D}\!\left(\mathbf{I} + \frac{1}{R_0^2}\,\mathbf{h}\mathbf{h}^\top\right),
\label{eq:A_inv}
\end{equation}
gives the 2-D bistatic signal model in the RD domain:
\begin{align}
s(t,\mathbf{k}_m \mid \mathbf{h}) &= A_0\,w_r\bigl(t - t(\mathbf{k}_m)\bigr) \nonumber \\
  &\quad\cdot \exp\!\left(-j\frac{4\pi f_0}{c}\sqrt{D}
  + j\frac{\pi c}{2f_0}\,\mathbf{k}_m^\top \mathbf{A}^{-1}\mathbf{k}_m\right),
\label{eq:signal_rd_2d} \\
t(\mathbf{k}_m) &= \frac{2\sqrt{D}}{c}
  + \frac{c}{4f_0^2}\,\mathbf{k}_m^\top \mathbf{A}^{-1}\mathbf{k}_m,
\label{eq:travel_time_rd_2d}
\end{align}
where $\mathbf{k}_m = (k_{m,x},\,k_{m,y})^\top \in \mathbb{R}^2$.
Equations~\eqref{eq:signal_rd_2d} - \eqref{eq:travel_time_rd_2d} mirror their 1-D counterparts \eqref{eq:signal_rd} - \eqref{eq:travel_time_rd}, with the scalar $K_m$ replaced by $\mathbf{A}$.
Beamforming proceeds by 2-D RCMC followed by 2-D matched filtering and a 2-D IFFT over $\mathbf{k}_m$.
Standard multistatic RDA on a 2-D aperture would require a 4-D FFT over all four aperture dimensions (two transmit, two receive) \cite{jakovljevic2022adaptation};
COG reduces this to a 2-D FFT per offset group, enabling volumetric frequency-domain beamforming of multistatic data within practical memory constraints.
The complete COG RDA processing pipeline, applicable to both 1-D and 2-D apertures, is summarized in Fig.~\ref{fig:cog_rda_flowchart}.

\subsection{Time and Space Complexity of COG RDA}
\label{subsec:complexity}

Algorithm complexity is described using the "big O" ($\mathcal{O}$) notation \cite{cormen2009introduction}, which we use here for two distinct quantities: \emph{time complexity}, the number of operations performed as a function of input and output array sizes (Sections~\ref{sssec:complexity_1d} and \ref{sssec:complexity_2d}), and \emph{space complexity}, the peak memory required to hold the largest data array during reconstruction (Section~\ref{sssec:complexity_memory}).
In both cases, only the dominant terms are kept.
The presented analysis is similar to our previous complexity analysis of monostatic and multistatic RDA in \cite{jakovljevic2022adaptation},
except that now, because all the discussed beamformers can reconstruct pixels/voxels at different depths independently,
we ignore the dependance on the number of time and depth samples ($T$ and $P$) and we express the complexity only in terms of the number of array elements $N$.
To simplify the resulting expressions, we assume that the number of reconstructed lateral pixels/voxels $M$ is of the same order as $N$.
We also assume that each bistatic gather is zero-padded to the same fixed size, so that every gather requires the same number of operations regardless of offset $h$ and the number of midpoints available at that specific offset.
That way, the total time complexity of a full COG reconstruction is simply the bistatic time complexity multiplied by the number of gathers.

\subsubsection{Time complexity for 1-D apertures}
\label{sssec:complexity_1d}

Time complexity of beamforming a single bistatic gather is identical to the monostatic case described in \cite{jakovljevic2022adaptation}.
The complexity of bistatic RDA is therefore
\begin{equation}
 \begin{array}{l}
\mathcal{O}\big(\underbracket[0.2ex][1ex]{\textstyle N\log_{2}N}_{\text{1-D FFT}} + \underbracket[0.2ex][1ex]{\textstyle M}_{\text{RCMC}} + \underbracket[0.2ex][1ex]{\textstyle M}_{\text{Matched filter}} + \underbracket[0.2ex][1ex]{\textstyle M\log_{2} M}_{\text{1-D IFFT}}\big) \sim \\ [4ex]
\mathcal{O}_{\text{RDA}}(2\,N\log_{2} N)\,,
\end{array}
\label{eq:complexity_bistatic_1d}
\end{equation}
with an equivalent bistatic DAS complexity of $\mathcal{O}_{\text{DAS}}(NM) \sim \mathcal{O}_{\text{DAS}}(N^2)$.

A full multistatic reconstruction requires beamforming all $N$ offset gathers.
Summing \eqref{eq:complexity_bistatic_1d} over $N$ diagonals gives the total COG RDA complexity,
\begin{equation}
\mathcal{O}_{\text{COG RDA}}(2\,N^2\log_{2} N)\,,
\label{eq:complexity_cog_multistatic_1d}
\end{equation}
which is the same order as standard multistatic RDA derived in \cite{jakovljevic2022adaptation}.
\footnote{Standard multistatic RDA has complexity $\mathcal{O}_{\text{RDA}}(2\,N^2\log_{2} N^2)$ \cite{jakovljevic2022adaptation},
a factor of 2 more operations than \eqref{eq:complexity_cog_multistatic_1d}.}
The corresponding full multistatic DAS complexity is $\mathcal{O}_{\text{DAS}}(N^3)$ \cite{jakovljevic2022adaptation}.

\subsubsection{Time complexity for 2-D apertures}
\label{sssec:complexity_2d}

For a fully sampled $N\times N$ matrix array used to reconstruct a $P\times M^2$ volume,
the (zero-padded) midpoint vector $\mathbf{m}$ of a single bistatic gather spans an $N\times N$ grid.
The complexity of volumetric bistatic RDA is
\begin{equation}
 \begin{array}{l}
\mathcal{O}\big(\underbracket[0.2ex][1ex]{\textstyle N^2\log_{2}N^2}_{\text{2-D FFT}} + \underbracket[0.2ex][1ex]{\textstyle M^2}_{\text{RCMC}} + \underbracket[0.2ex][1ex]{\textstyle M^2}_{\text{Matched filter}} + \underbracket[0.2ex][1ex]{\textstyle M^2\log_{2} M^2}_{\text{2-D IFFT}}\big) \sim \\ [4ex]
\mathcal{O}_{\text{RDA}}(2\,N^2\log_{2} N^2)\,.
\end{array}
\label{eq:complexity_bistatic_2d}
\end{equation}
The complexity of bistatic volumetric DAS is given by $\mathcal{O}_{\text{DAS}}(N^2M^2) \sim \mathcal{O}_{\text{DAS}}(N^4)$.

A full volumetric reconstruction requires beamforming all $N^2$ offset gathers (one gather for each 2-D offset vector $\mathbf{h}$).
Summing \eqref{eq:complexity_bistatic_2d} over $N^2$ gathers gives the total COG RDA complexity,
\begin{equation}
\mathcal{O}_{\text{COG RDA}}(2\,N^4\log_{2} N^2)\,.
\label{eq:complexity_cog_multistatic_2d}
\end{equation}
Computational complexity of volumetric multistatic DAS is $\mathcal{O}_{\text{DAS}}(N^2 \cdot N^2M^2) \sim \mathcal{O}_{\text{DAS}}(N^6)$.

\subsubsection{Space complexity}
\label{sssec:complexity_memory}
Space complexity measures how much memory algorithms require as the input size $N$ increases.
For a 1-D array, standard multistatic RDA takes a 2-D FFT of the data matrix along the Tx and Rx dimensions, requiring memory on the order of $\mathcal{O}(N^2)$.
COG RDA instead takes a 1-D FFT along midpoints $m$, and holds only a single diagonal of the data matrix at a time, giving the space complexity of $\mathcal{O}(N)$.
For an $N\times N$ matrix array, standard volumetric RDA would require the full 4-D range-Doppler tensor, $\mathcal{O}(N^4)$.
Volumetric COG RDA instead holds only a single 2-D offset gather, $\mathcal{O}(N^2)$.

As a reference, pixel-based DAS accumulates the delayed individual channels one pixel at a time, requiring only enough memory to store a single trace, $\mathcal{O}(1)$.

Time and space complexity for DAS, standard multistatic RDA, and COG RDA are summarized for 1-D and 2-D apertures in Table~\ref{tab:complexity_summary}.
\begin{table}[!t]
\caption{Time and space complexity of DAS, standard multistatic RDA, and COG RDA for 1-D and 2-D apertures,
expressed as a function of the number of elements $N$ per aperture dimension.}
\label{tab:complexity_summary}
\centering
\footnotesize
\setlength{\tabcolsep}{5pt}
\begin{tabular}{@{}lcccc@{}}
\hhline{=====}
& \multicolumn{2}{c}{1-D Aperture} & \multicolumn{2}{c}{2-D Aperture} \\
Method & Time & Space & Time & Space \\
\hhline{=====}
DAS & $\mathcal{O}(N^3)$ & $\mathcal{O}(1)$ & $\mathcal{O}(N^6)$ & $\mathcal{O}(1)$ \\
Standard RDA & $\mathcal{O}(2N^2\log_{2} N^2)$ & $\mathcal{O}(N^2)$ & $\mathcal{O}(2N^4\log_{2} N^4)$ & $\mathcal{O}(N^4)$ \\
COG RDA & $\mathcal{O}(2N^2\log_{2} N)$ & $\mathcal{O}(N)$ & $\mathcal{O}(2N^4\log_{2} N^2)$ & $\mathcal{O}(N^2)$ \\
\hhline{=====}
\end{tabular}
\end{table}

\section{Methods}
\label{sec:methods}
\subsection{FIELD II Simulations}
\label{subsec:simulations}

\subsubsection{1-D Linear Array}
\label{ssubsec:sim_1d}

Multistatic synthetic aperture (SA) channel data were simulated using Field~II \cite{Jensen96} for a 128-element 1-D linear array.
Simulated phantom contained point targets placed in 10\,mm increments in $z$ dimension to assess lateral resolution as a function of depth.
Array and acquisition parameters are listed in Table~\ref{tab:acq_params}.

\begin{table}[!t]
\caption{Array and acquisition parameters for the simulations and for the phantom and \emph{in vivo} acquisitions.}
\label{tab:acq_params}
\centering
\footnotesize
\setlength{\tabcolsep}{3pt}
\renewcommand{\arraystretch}{1.1}
\begin{tabularx}{\columnwidth}{@{}l*{3}{>{\raggedright\arraybackslash}X}@{}}
\hhline{====}
                     & 1-D simulation & Matrix simulation & Phantom / \emph{in vivo} \\
\hhline{----}
Array                & 128-element linear & $32 \times 32$ matrix & GE L2-9 linear (192 elements) \\
Element pitch        & 0.15\,mm & 0.5\,mm & 0.23\,mm \\
Aperture size        & 19.2\,mm & $16 \times 16$\,mm & 44\,mm \\
Center frequency     & 5\,MHz & 2.5\,MHz & 5.5\,MHz \\
Fractional bandwidth & 60\,\% & 60\,\% & -- \\
Sampling frequency   & -- & 10\,MHz & -- \\
Speed of sound       & 1540\,m/s & 1540\,m/s & -- \\
Transmit sequence    & Full SA (128 single-element) & Explososcan ($22 \times 15 = 330$ focused beams) & Linear scan (80 focused beams) \\
Scanning angles      & -- & \degrees{30} az.\ $\times$ \degrees{20} el. & -- \\
Tx focal depth       & -- & 50\,mm & 120\,mm \\
System               & Field~II & Field~II & GE LOGIQ E10 \\
\hhline{====}
\end{tabularx}
\end{table}

\subsubsection{Matrix Arrays}
\label{ssubsec:sim_matrix}

Field~II \cite{Jensen96} was also used to simulate pre-beamformed ultrasound signals from a fully sampled, $32 \times 32$-element matrix array, acquired with an Explososcan transmit sequence \cite{Shattuck84,vonRamm91b}.
Explososcan uses a set of broadly focused transmit beams that span the azimuth and elevation dimensions to cover a wide field of view, and was originally proposed on early 3-D systems to insonify large volumes at real-time acquisition rates.
A 3-D point target phantom was simulated to evaluate beamformer resolution as a function of depth and steering angle;
targets were distributed over a depth range of 10 -- 70\,mm and placed as far as $\pm$15\,mm in lateral and elevation dimensions to evaluate image quality beyond the aperture footprint.
Volumetric simulation parameters are listed in Table~\ref{tab:acq_params}.

\subsection{Phantom Acquisitions}
\label{subsec:phantom}

Pre-beamformed receive channel data was collected with a L2-9 linear array connected to a GE LOGIQ E10 clinical ultrasound scanner (GE HealthCare, Waukesha, WI)
while scanning a CIRS calibrated tissue-mimicking phantom (CIRS Model 040GSE, Norfolk, VA).
Regions containing closely spaced wire targets and hyperechoic cylindrical lesions surrounded by diffuse scatterers were scanned using
a focused transmit sequence with the acquisition parameters listed in Table~\ref{tab:acq_params}.

Full synthetic aperture (SA) data was recovered from the focused transmit sequence using the REFoCUS method \cite{bottenus2017recovery,ali2019extending}.
REFoCUS uses known Tx delay and apodization profiles to invert the transmit focusing matrix and defocus the signals after the acquisition,
effectively estiamting the full SA dataset that would have been acquired with single-element transmits.

\subsection{In Vivo Acquisitions}
\label{subsec:invivo}

The \emph{in vivo} study was in accordance with the Health Insurance Portability and Accountability Act and was approved by the Institutional Review Board (IRB) of Massachusetts General Hospital under protocol number 2024P000693.
Verbal consent was obtained from the participant prior to enrollment.
Pre-beamformed receive channel data were acquired during an abdominal scan on a healthy volunteer,
using the same scanner, transducer, and focused transmit sequence as in the phantom experiment (GE LOGIQ E10 with the L2-9 linear array; Table~\ref{tab:acq_params}).
The acquisition was performed by a trained radiologist, who scanned the upper abdomen in a transverse plane to capture liver tissue and vasculature.
As for the phantom data, the full multistatic dataset was recovered from the focused-transmit acquisition using REFoCUS \cite{bottenus2017recovery,ali2019extending}.

\subsection{Image Reconstruction}
\label{subsec:reconstruction}

All images were reconstructed offline.
For 1-D arrays, COG RDA was compared with multistatic delay-and-sum (DAS) \cite{Cobbold07} and standard multistatic RDA \cite{jakovljevic2022adaptation}.
For matrix arrays, it was compared with 3-D DAS only, since standard multistatic RDA would require an impractical 4-D FFT over all four Tx/Rx aperture dimensions.
For the phantom 2-D scans and the simulated volumetric acquisition, multistatic data were first recovered from the focused-transmit data using REFoCUS \cite{bottenus2017recovery,ali2019extending}.
For the \emph{in vivo} acquisition, COG RDA (on REFoCUS-recovered data) was compared with DAS RTB, the manufacturer-optimized clinical reference, and with RDA RTB \cite{jakovljevic2026power}, both applied directly to the focused-transmit data.

Each constant-offset diagonal was beamformed independently and the partial images/volumes were accumulated coherently (Section~\ref{sec:theory}).
The matched filter for each offset was derived analytically from Eqs. \eqref{eq:signal_rd} and \eqref{eq:travel_time_rd} for linear arrays, and Eqs. \eqref{eq:signal_rd_2d} and \eqref{eq:travel_time_rd_2d} for matrix arrays.
RCMC was performed by 1-D linear interpolation along the range axis, applied at each $k_m$ ($\mathbf{k}_m$) sample.

All 2-D beamformers were implemented in MATLAB (MathWorks, Natick, MA) and executed on a single CPU core.
Our MATLAB code for 1-D COG RDA and the simulation example in Fig.~\ref{fig:simulated_2d_results} is available in the existing repository {\color{blue}\url{https://gitlab.com/mj66/bentobox}}.
Volumetric DAS used the open-source \texttt{rtbf} GPU beamforming library \cite{hyun2019open}, applied to the demodulated (IQ) channel data.
Volumetric COG RDA was implemented in Python with CuPy \cite{cupy2025} and a custom CUDA kernel for the RCMC interpolation.
Both ran on a laptop-based NVIDIA GeForce RTX 3080 GPU (16\,GB of video memory, 80\,W maximum power draw).

\subsection{Performance Metrics}
\label{subsec:metrics}

\subsubsection{Point Spread Functions (PSFs)}
For 2-D images from linear arrays, lateral resolution was quantified from point target images as the full width at half maximum (FWHM) of the lateral intensity profile at each target depth.
The mainlobe-to-sidelobe energy ratio (MSER) was also computed as the ratio of energy inside the mainlobe region ($> -6$\,dB) to the sidelobe energy between $-6$\,dB and $-50$\,dB relative to the PSF peak.
For 3-D volumes, the PSF metrics included the mainlobe volume (within the $-6$\,dB bounds) and the MSER.

\subsubsection{Contrast}
Image contrast was quantified using the generalized contrast-to-noise ratio, $\mathrm{gCNR} = 1 - \mathrm{OVL}$ \cite{Rodriguez19generalized}, where OVL is the overlap between the probability density functions of pixel intensities inside the lesion and in the surrounding background.
Both densities were approximated by histograms of log-compressed pixel values (100 bins from $-50$ to $0$\,dB).

\subsubsection{In Vivo Image Quality Metrics}
To compare the overall appearance of the \emph{in vivo} images to the DAS RTB clinical reference, the structural similarity index (SSIM)~\cite{wang2004image} was computed from log-compressed images interpolated onto the COG RDA grid.
Each image was first smoothed with a 1\,mm Gaussian kernel, so that the comparison reflects tissue structure rather than individual speckle agreement.

\subsubsection{Beamforming Efficiency}
Beamforming time was measured from raw channel data to the envelope-detected image, and memory as the size of the largest intermediate array allocated during reconstruction.
For volumetric reconstructions, GPU energy, power, utilization, and temperature increase were also recorded.

\section{Results}
\label{sec:results}
\subsection{Pulse-echo Path Approximation}
\label{subsec:results_approximation}

\begin{figure}[!t]
\centering
\includegraphics[width=\columnwidth,trim={0 0 1cm 0},clip]{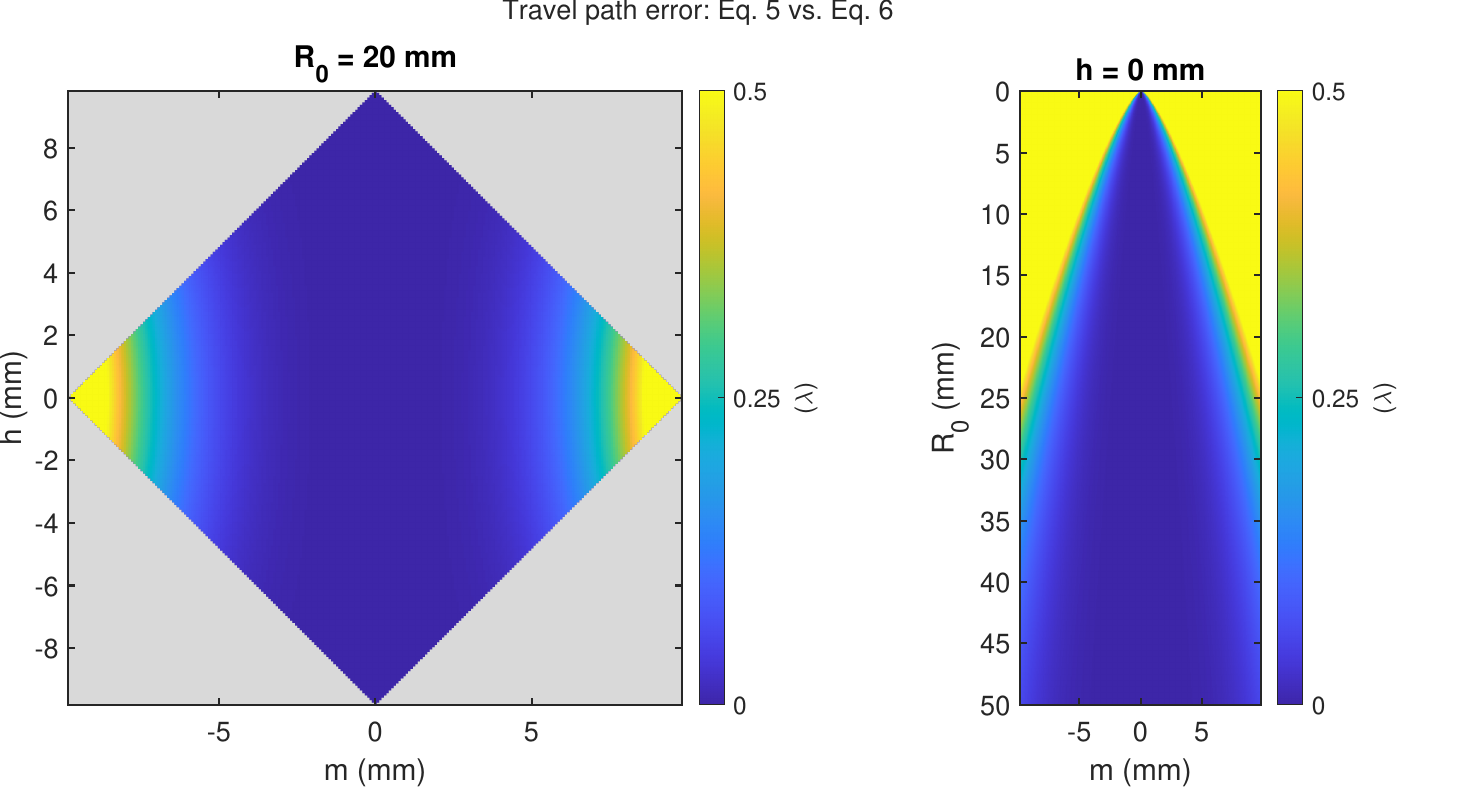}
\caption{Phase error of the second-order travel-path approximation in Eq. \eqref{eq:travel_path_1d} relative to the exact bistatic travel distances, expressed in wavelengths~$\lambda$.
(a)~Error as a function of midpoint $m$ and half-offset $h$ at a fixed scatterer depth $R_0 = 20$~mm;
the gray region indicates element pairs outside the physical aperture ($|m| + |h| >$ array half-aperture).
(b)~Error along the zero-offset diagonal ($h = 0$) as a function of midpoint $m$ and scatterer depth $R_0$.
The approximation error remains well below $0.25\,\lambda$ for depths larger than 30\,mm, validating the second-order travel-path model for the COG RDA signal in \eqref{eq:signal_rd}.}
\label{fig:travel_path_error}
\end{figure}

Fig.~\ref{fig:travel_path_error} evaluates the accuracy of the second-order approximation for pulse-echo travel-path in Eq. \eqref{eq:travel_path_1d}
by computing the error relative to the exact bistatic travel distances in Eq. \eqref{eq:travel_path_1d_complete}.
The error is expressed in wavelengths and computed across the FOV.
Fig.~\ref{fig:travel_path_error}(a) shows that, at a fixed depth of $R_0 = 20$\,mm, the error is largest along the zero-offset diagonal ($h = 0$) as the midpoint $m$ approaches the edge of the physical aperture, and decreases as $|h|$ increases at a fixed total aperture extent $|m| + |h|$.
Fig.~\ref{fig:travel_path_error}(b) shows that (along the zero-offset diagonal) model error is substantial for f-numbers $\leq 1$, where it exceeds $0.5\lambda$,
Beyond $\sim$30\,mm depth, the error remains well below $0.25\,\lambda$,
confirming the suitability of the second-order approximation for the COG RDA signal models in \eqref{eq:signal_rd} and \eqref{eq:signal_rd_2d}.

\subsection{2-D B-mode Imaging}
\label{subsec:results_2d}

\begin{figure}[!t]
\centering
\begin{tabular}{@{}c@{}}
\includegraphics[width=0.95\columnwidth]{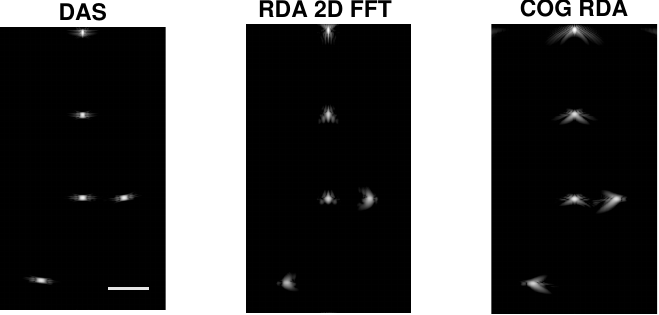}\\[0.5ex]
\textbf{(a)}\\[1ex]
\includegraphics[width=0.82\columnwidth]{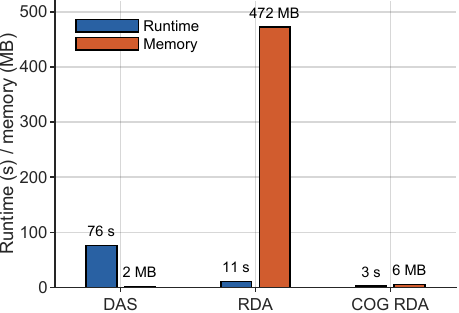}\\[0.5ex]
\textbf{(b)} \\[0.5ex]
\end{tabular}
\caption{Simulated 2-D point-target results.
(a)~B-mode images reconstructed from the multistatic Field II data using DAS, conventional RDA, and COG RDA.
(b)~Runtime and maximum variable size for each reconstruction methods, indicating their respective computational and memory requirements.}
\label{fig:simulated_2d_results}
\end{figure}

Fig.~\ref{fig:simulated_2d_results}(a) compares the simulated, 2-D point-target images reconstructed using DAS, conventional multistatic RDA, and the proposed COG RDA implementation.
All images are normalized by the maximum, log compressed, and displayed over 50\,dB of dynamic range.
The scale bar shown in the DAS image is 5\,mm.
The on-axis targets ($x=0$, at $z=10$, $20$, and $30$\,mm) are visually comparable across all three methods, each retaining a compact mainlobe and low sidelobes.
The off-axis targets (located at (x,z) coordinates (5,\,30) and (-5,\,40)\,mm) show a wedge-shaped sidelobe artifact in the RDA and COG RDA images that is not present in the DAS image.

Table~\ref{tab:simulation_pt_target_metrics_2d} lists the lateral full-width at half-maximum (FWHM$_x$) and mainlobe-to-sidelobe energy ratio (MSER) measured from the point targets in Fig.~\ref{fig:simulated_2d_results}(a).
\begin{table}[!t]
    \caption{Simulated 2-D point-target resolution and sidelobe metrics.}
    \label{tab:simulation_pt_target_metrics_2d}
    \centering
    \renewcommand{\arraystretch}{1.15}
    \footnotesize
    \setlength{\tabcolsep}{4pt}
    \begin{tabular}{cccccccc}
    \hline
    \multicolumn{2}{c}{Target pos.} & \multicolumn{2}{c}{DAS} & \multicolumn{2}{c}{RDA 2D FFT} & \multicolumn{2}{c}{COG RDA} \\
    $x$ & $z$ & FWHM$_x$ & MSER & FWHM$_x$ & MSER & FWHM$_x$ & MSER \\
    (mm) & (mm) & (mm) & (dB) & (mm) & (dB) & (mm) & (dB) \\
    \hline
    0 & 10 & 0.2 & 5.0 & 0.2 & 5.3 & 0.4 & 4.3 \\
0 & 20 & 0.3 & 5.1 & 0.4 & 4.5 & 0.5 & 3.3 \\
5 & 30 & 0.5 & 5.2 & 0.5 & 2.7 & 0.6 & 2.5 \\
0 & 30 & 0.5 & 5.6 & 0.5 & 4.6 & 0.6 & 4.1 \\
-5 & 40 & 0.6 & 5.2 & 0.6 & 2.9 & 0.7 & 3.5 \\
\hline
\hline
    \end{tabular}
\end{table}

FWHM$_x$ grows almost linearly with depth for all three methods, as expected,
with COG RDA showing a slight broadening ($\leq 0.2$\,mm) relative to DAS and conventional RDA at each target location.
On axis, conventional RDA preserves the MSER of DAS more closely than COG RDA.
Both frequency-domain methods show a further, substantial MSER reduction at the off-axis target locations,
which is consistent with the wedge-shaped sidelobe artifacts observed in Fig.~\ref{fig:simulated_2d_results}(a).
Although the artifact is visibly larger for COG RDA than for conventional RDA at both off-axis locations,
the corresponding MSER values are comparable between the two methods.

Fig.~\ref{fig:simulated_2d_results}(b) reports the runtime and memory requirements for the image reconstructions in Fig.~\ref{fig:simulated_2d_results}(a).
COG RDA is the fastest method, with a $3.7\times$ and $25\times$ speedup relative to conventional RDA and DAS, respectively.
COG RDA also requires $79\times$ less memory than conventional RDA, which stores the channel data in the range-Doppler domain after a 2-D FFT over the full Tx/Rx aperture,
whereas COG RDA holds only a single-offset diagonal in memory at a time.
DAS, which need not store the full channel data array simultaneously, requires $3\times$ less memory than COG RDA.

\begin{figure}[!t]
\centering
\begin{tabular}{@{}c@{}}
\includegraphics[width=0.95\columnwidth]{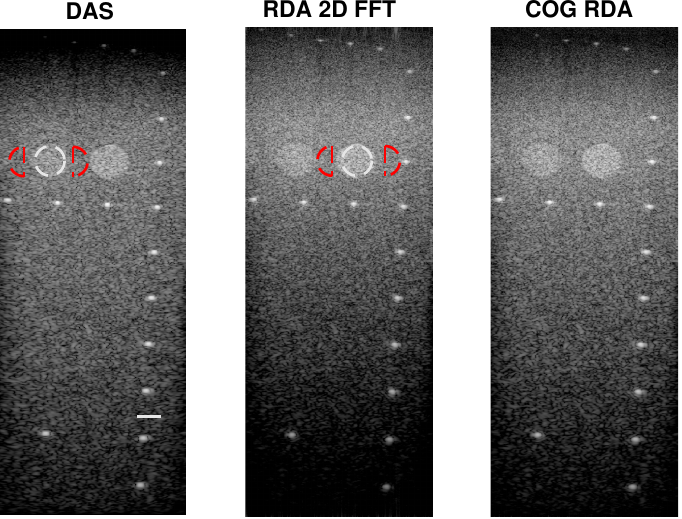}\\[0.5ex]
\textbf{(a)}\\[1ex]
\includegraphics[width=0.82\columnwidth]{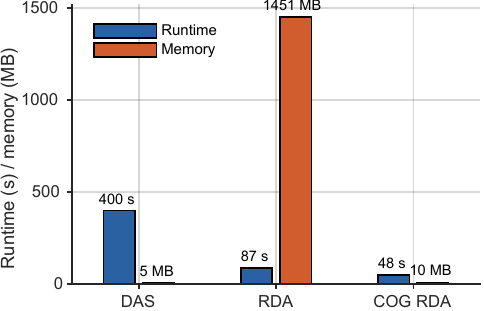}\\[0.5ex]
\textbf{(b)} \\[0.5ex]
\end{tabular}
\caption{2-D B-mode results from the CIRS calibrated tissue-mimicking phantom.
(a)~B-mode images reconstructed using DAS, conventional RDA, and COG RDA.
(b)~Runtime and maximum variable size for each reconstruction methods.}
\label{fig:phantom_2d_results}
\end{figure}

Fig.~\ref{fig:phantom_2d_results}(a) compares the same three multistatic reconstruction methods on the CIRS phantom acquisition, with the multistatic data recovered from the focused-transmit acquisition using the REFoCUS method.
The wire targets and the two hyperechoic lesions near the top of the field of view appear qualitatively similar across DAS, conventional multistatic RDA, and COG RDA.
Dashed circles in Fig.~\ref{fig:phantom_2d_results}(a) denote the regions of interest used to compute the gCNR values for the two lesions in Table~\ref{tab:phantom_lesion_gcnr}.
gCNR is computed from these regions for all three beamforming methods, but the overlay is shown only on a subset of panels to avoid cluttering the B-mode images and to preserve image clarity.

\begin{table}[!t]
    \caption{Image quality metrics for the CIRS phantom acquisition.}
    \label{tab:phantom_lesion_gcnr}
    \centering
    \renewcommand{\arraystretch}{1.15}
    \small
    \setlength{\tabcolsep}{6pt}
    \begin{tabular}{cccc}
    \hline
    Metric & DAS & RDA 2D FFT & COG RDA \\
    \hline
    Left lesion gCNR & 0.52 & 0.53 & 0.52 \\
Right lesion gCNR & 0.64 & 0.61 & 0.63 \\
Avg. pt. target area & / & 18.6 & 6.0 \\
diff. vs DAS (\%) & & & \\
\hline
\hline
    \end{tabular}
\end{table}

Table~\ref{tab:phantom_lesion_gcnr} summarizes the phantom image quality metrics for Fig.~\ref{fig:phantom_2d_results}(a).
gCNR values are nearly identical across methods for both lesions, indicating that lesion contrast is preserved by both frequency-domain methods relative to DAS.
The average point-target mainlobe area is larger for the frequency-domain methods than for DAS, consistent with the modest resolution loss observed for the simulated point targets in Table~\ref{tab:simulation_pt_target_metrics_2d}.

Barplots in Fig.~\ref{fig:phantom_2d_results}(b) report the runtime and memory requirements for the phantom reconstructions.
As in the simulated case, COG RDA is the fastest method, achieving a $1.8\times$ speedup relative to standard RDA and $8.3\times$ speedup relative to DAS.
COG RDA also reduces memory requirements $145\times$ relative to conventional multistatic RDA, while remaining only $2\times$ above multistatic DAS.
The larger absolute runtimes and memory values relative to the simulated case reflect the larger acquisition depth (10\,cm for the phantom acquisition compared to 6\,cm for the simulated scan),
and higher element count of the $192$-element L2-9 array compared to the $128$-element simulated array.

\begin{figure}[!t]
\centering
\includegraphics[width=0.95\columnwidth]{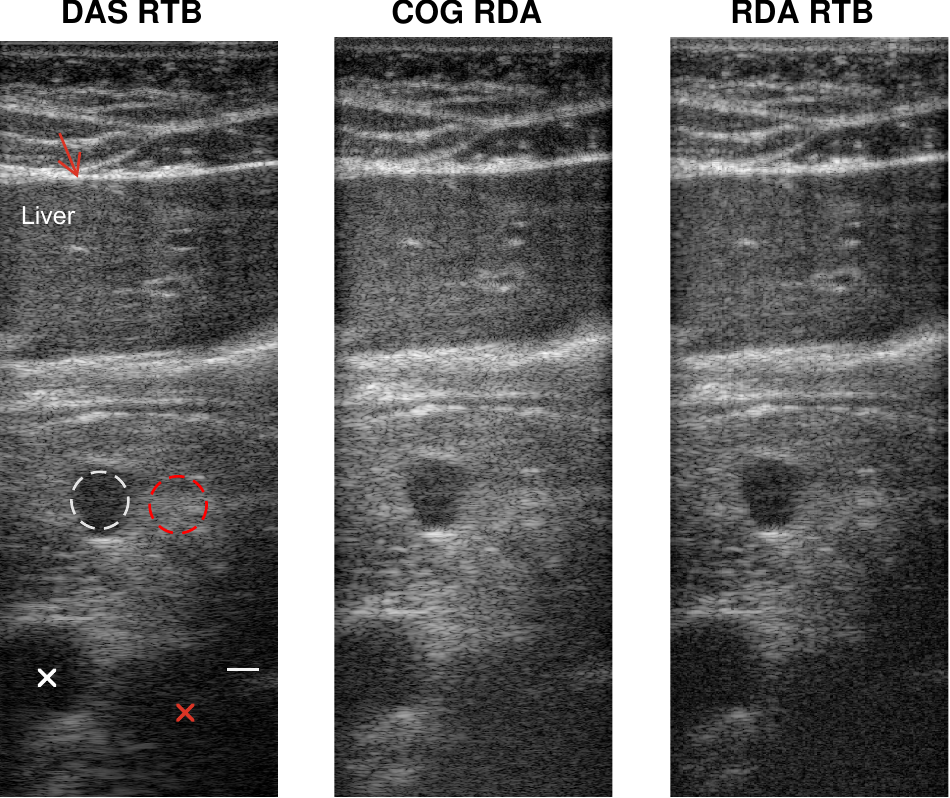}
\caption{\emph{In vivo} B-mode images of the mid-upper abdomen of a healthy volunteer (transverse plane), reconstructed using DAS RTB (as optimized by the scanner manufacturer) and RDA RTB applied to the focused-transmit data, and COG RDA applied to the multistatic data recovered with REFoCUS.
In the DAS RTB image, the white and red dashed circles mark the origin of the portal vein and the echogenic perivascular tissue (gCNR regions), the white and red crosses mark the inferior vena cava and the abdominal aorta, and the red arrow indicates the liver capsule.
All images are displayed over 60\,dB of dynamic range; the scale bar is 5\,mm.}
\label{fig:invivo_results}
\end{figure}

Fig.~\ref{fig:invivo_results} shows B-mode images from the \emph{in vivo} abdominal acquisition, reconstructed using DAS RTB (as clinical reference), COG RDA, and RDA RTB.
Time-gain compensation (TGC) was applied to the COG RDA and RDA RTB images to adjust brightness at larger depths, and all three images are displayed over a 60\,dB dynamic range.
The abdominal wall layers, the liver, and the major abdominal vessels are visualized at the same locations and with similar contrast in all three images.
Table~\ref{tab:invivo_metrics} summarizes gCNR of the portal vein relative to the surrounding perivascular tissue, structural similarity (SSIM) relative to DAS RTB, and beamforming runtime.

\begin{table}[!t]
    \caption{Image quality metrics and CPU beamforming time for the \emph{in vivo} acquisition in Fig.~\ref{fig:invivo_results}.
    Regions used to compute gCNR are denoted in the DAS RTB B-mode image.
    All timings are measured in MATLAB and exclude data loading and display.}
    \label{tab:invivo_metrics}
    \centering
    \renewcommand{\arraystretch}{1.15}
    \small
    \setlength{\tabcolsep}{6pt}
    \begin{tabular}{cccc}
    \hline
    Metric & DAS RTB & COG RDA & RDA RTB \\
    \hline
    gCNR & 0.68 & 0.64 & 0.64 \\
SSIM (vs. DAS RTB) & 1.00 & 0.74 & 0.77 \\
Runtime (s) & 212 & 55 & 141 \\
\hline
\hline
    \end{tabular}
\end{table}

\subsection{Volumetric Imaging}
\label{subsec:results_volumetric}

\begin{figure}[!t]
\centering
\setlength{\tabcolsep}{1pt}
\renewcommand{\arraystretch}{0.95}
\begin{tabular}{@{}c@{\hspace{2pt}}c@{}}
\includegraphics[width=0.31\columnwidth,trim={0.85cm 1.2cm 1.2cm 1.0cm},clip]{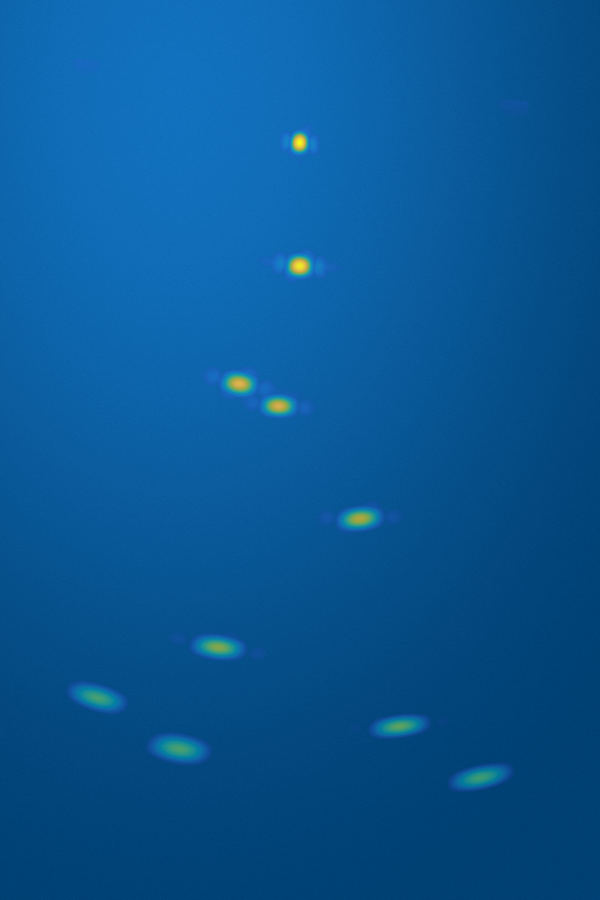} &
\includegraphics[width=0.66\columnwidth,trim={0cm 0cm 0cm 0cm},clip]{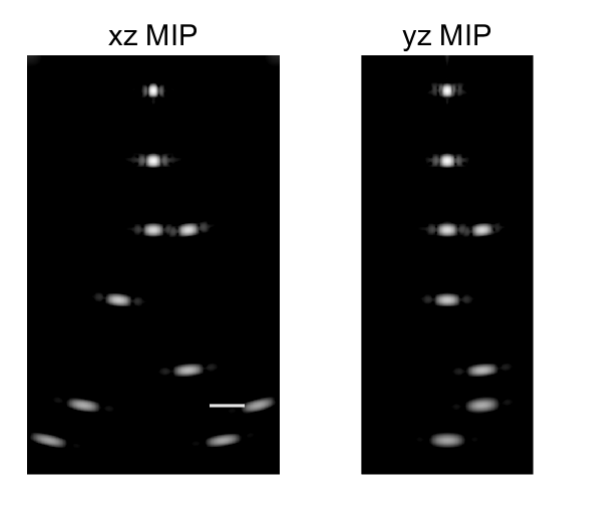} \\[-0.5ex]
\multicolumn{2}{@{}c@{}}{\textbf{(a) \small 3-D DAS}} \\[0.75ex]
\includegraphics[width=0.31\columnwidth,trim={0.85cm 1.2cm 1.2cm 1.0cm},clip]{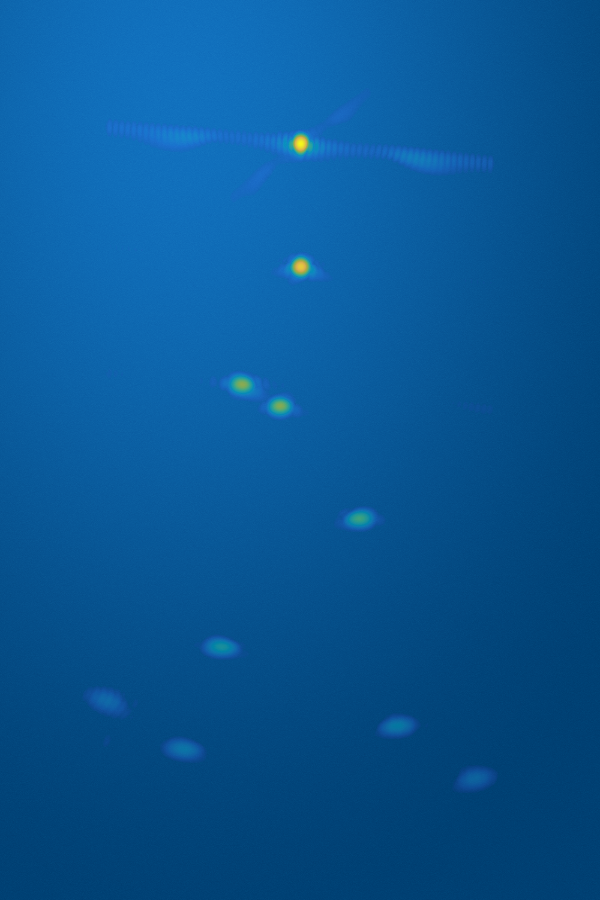} &
\includegraphics[width=0.66\columnwidth]{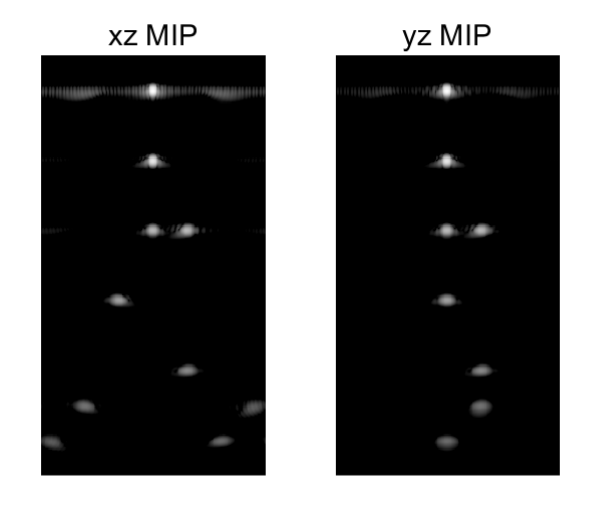} \\[-0.5ex]
\multicolumn{2}{@{}c@{}}{\textbf{(b) \small 3-D COG RDA}} \\[0.75ex]
\end{tabular}
\caption{DAS and COG RDA volumetric reconstructions of the simulated 3-D sector scan (Explososcan).
Each row shows the volume rendering and corresponding $xz$ and $yz$ maximum-intensity projections (MIP) for one beamformer: (a)~3-D DAS (GPU) and (b)~3-D COG RDA.
All images are displayed on the decibel scale from $-50$ to $0$\,dB; the scale bar in the $xz$ MIPs is 5\,mm.}
\label{fig:3d_vols_exploso}
\vspace{0.2cm}
\centering
\includegraphics[width=0.9\columnwidth]{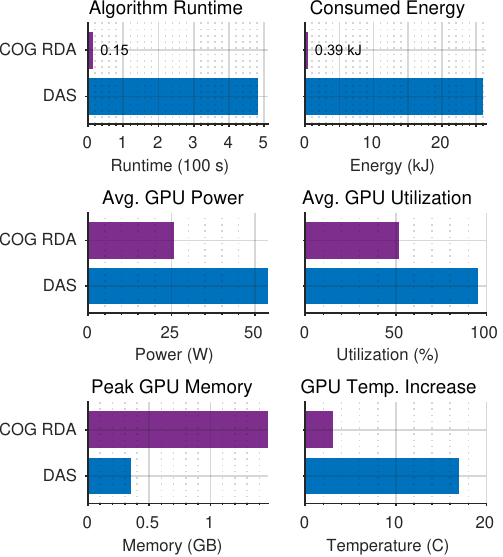} \\[0.5ex]
\caption{GPU consumption metrics for a single reconstructed volume using 3-D DAS and 3-D COG RDA. The metrics include algorithm runtime, total energy consumed,
average power and percent utilization, peak memory and temperature increase.
Peak memory is reported relative to the idle-before GPU memory baseline.}
\label{fig:3d_gpu_resource_bars}
\end{figure}

Fig.~\ref{fig:3d_vols_exploso} compares the DAS and COG RDA volumes reconstructed from the simulated 3-D sector scan (Explososcan), with corresponding maximum-intensity projections (MIPs) shown for each method.
All images are displayed over a 50\,dB dynamic range; scale bars in the $xz$ MIPs are 5\,mm.
The 3-D point targets appear visually comparable between the two beamforming approaches across the full reconstructed volume.
Both beamformers are able to clearly visualize targets beyond the extent of the physical aperture.
On the same GPU, COG RDA beamforms the volume approximately $32\times$ faster than DAS (Fig.~\ref{fig:3d_gpu_resource_bars}).

Table~\ref{tab:psf_metrics} lists the 3-D point target PSF volume and MSER for the reconstructions in Fig.~\ref{fig:3d_vols_exploso}.
PSF volume increases with depth for both methods and is nearly identical between them down to 60\,mm depth; only at the deepest target (65\,mm) is the COG RDA mainlobe volume noticeably larger.
COG RDA yields an MSER equal to or higher than that of DAS at every target location, confirming that the frequency-domain method preserves sidelobe suppression relative to the GPU-based DAS baseline.

Fig.~\ref{fig:3d_gpu_resource_bars} summarizes the GPU resource consumption for each beamformer during reconstruction of a single volume.
3-D DAS consumes approximately 26\,kJ of energy to reconstruct a volume, while maintaining near-full GPU utilization (95\%) at an average power draw of about 54\,W throughout the reconstruction.
COG RDA requires substantially less energy to reconstruct a volume (0.39\,kJ, a $67\times$ reduction relative to DAS);
it also achieves lower average GPU utilization (52\%) and power draw (26\,W).
Despite processing the data sequentially one offset at a time, COG RDA incurs a larger peak GPU memory increase above the idle baseline (1.5\,GB vs.\ 0.35\,GB for DAS),
reflecting the additional memory needed to stage the reconstructed volume and intermediate RCMC interpolation buffers on the GPU.
The lower average computational load of COG RDA also results in a smaller GPU temperature increase during reconstruction (3\,\textdegree{}C vs.\ 17\,\textdegree{}C for DAS).

\begin{table}[!t]
\caption{3-D point target PSF volume and MSER for the Explososcan reconstructions in Fig.~\ref{fig:3d_vols_exploso}.}
\label{tab:psf_metrics}
\centering
\small
\renewcommand{\arraystretch}{1.1}
\setlength{\tabcolsep}{4pt}
\begin{tabularx}{\columnwidth}{*{3}{>{\centering\arraybackslash}p{0.6cm}}@{\hspace{4pt}}|@{\hspace{4pt}}*{4}{>{\centering\arraybackslash}X}}
\hline
\multicolumn{3}{c}{Target pos. (mm)} & \multicolumn{2}{c}{Mainlobe vol. (mm$^3$)} & \multicolumn{2}{c}{MSER} \\
$x$ & $y$ & $z$ & DAS & \mbox{COG RDA} & DAS & \mbox{COG RDA} \\
\hline
0 & 0 & 15 & 0.2 & 0.3 & 1.39 & 1.45 \\
0 & 0 & 25 & 0.4 & 0.4 & 1.37 & 1.50 \\
0 & 5 & 35 & 0.8 & 0.8 & 1.37 & 1.57 \\
-5 & 0 & 45 & 1.2 & 1.1 & 1.39 & 1.50 \\
5 & 5 & 55 & 1.8 & 1.7 & 1.40 & 1.65 \\
-10 & 5 & 60 & 2.4 & 2.4 & 1.49 & 1.49 \\
-15 & 0 & 65 & 2.8 & 4.2 & 1.46 & 1.99 \\
\hline
\hline
\end{tabularx}
\end{table}

\section{Discussion}
\label{sec:discussion}
\subsection{Image Quality Relative to DAS and Conventional Multistatic RDA}

For 1-D linear arrays, COG RDA preserves image quality relative to both DAS and conventional multistatic RDA, with only a slight lateral broadening and a slightly larger on-axis MSER loss (Table~\ref{tab:simulation_pt_target_metrics_2d}),
consistent with the small error of the second-order travel-path approximation (Fig.~\ref{fig:travel_path_error}).
Both frequency-domain methods show a larger MSER loss and wedge-shaped sidelobe artifacts at the off-axis targets (Fig.~\ref{fig:simulated_2d_results}(a)).
Because conventional RDA also shows this artifact despite relying on the exact travel-path model \cite{jakovljevic2022adaptation}, the narrowband assumption common to both RDA beamformers is its primary source.
The more pronounced artifact in COG RDA is consistent with the approximation error, which is largest near the edges of the physical aperture (Fig.~\ref{fig:travel_path_error}(a)).

The \emph{in vivo} images (Fig.~\ref{fig:invivo_results}) show that COG RDA preserves the main tissue layers, interfaces, and hypoechoic structures relative to DAS RTB and RDA RTB, with the same vessel gCNR as RDA RTB but lower SSIM (Table~\ref{tab:invivo_metrics}).
The coarser speckle at depth in both RDA images is consistent with their shared narrowband assumption:
frequency-dependent attenuation lowers the center frequency of the received echoes with depth, while the RDA matched filter assumes a constant center frequency throughout the FOV \cite{jakovljevic2026power}.
Depth-dependent matched filters could address this without affecting the compute efficiency or per-offset memory footprint of COG RDA.

For matrix arrays, COG RDA was compared with 3-D DAS only (Section~\ref{subsec:reconstruction}).
The volumes in Fig.~\ref{fig:3d_vols_exploso} show that the image-quality trends observed for 1-D arrays extend to volumetric imaging:
point targets are visually comparable, including off-axis targets beyond the physical extent of the aperture;
mainlobe volumes are nearly identical except at the deepest target; and COG RDA yields MSER values equal to or higher than those of DAS at all depths (Table~\ref{tab:psf_metrics}).
Only the shallowest target (at $15$\,mm depth) shows more pronounced lateral sidelobes in the COG RDA volume (Fig.~\ref{fig:3d_vols_exploso}(b)),
consistent with a larger second-order-approximation error at low f-numbers and with aperture undersampling at shallow depths.
These sidelobes extend beyond the region used to compute the PSF metrics and are therefore not captured by MSER.

\subsection{Memory and Runtime Efficiency from Independent Offset-Gather Processing}

While COG RDA and conventional multistatic RDA share the same asymptotic time complexity for 1-D apertures (Section~\ref{sssec:complexity_1d}),\footnote{Conventional RDA requires roughly twice as many operations, since its joint 2-D FFT scales with $\log_2 N^2$ rather than the $\log_2 N$ of the $N$ independent 1-D FFTs performed by COG.}
COG RDA has a factor of $N$ reduced space complexity (Section~\ref{sssec:complexity_memory}), leading to significantly lower memory requirements and reduced beamforming times in practice.
The two-orders-of-magnitude memory reductions relative to conventional RDA measured for both the simulated and phantom data (Figs.~\ref{fig:simulated_2d_results}(b) and \ref{fig:phantom_2d_results}(b))
are consistent with the $\mathcal{O}(N)$ versus $\mathcal{O}(N^2)$ space complexity listed in Table~\ref{tab:complexity_summary}.

The factor of $2$--$3$ gap in measured memory footprint between DAS and COG RDA is much smaller than the theoretical gap of $\mathcal{O}(1)$ versus $\mathcal{O}(N)$ predicted between the idealized versions of the two beamformers (Table \ref{tab:complexity_summary}).
The reason is that the DAS beamformer used here processes all $N$ receive-channel signals from a single transmit together, rather than one channel at a time.
Its practical memory footprint is therefore also on the order of $\mathcal{O}(N)$, similar to the single offset diagonal held by COG RDA, which explains their comparable measured memory use.
This combination of runtime and memory efficiency allows COG RDA to approach the low memory footprint of DAS while retaining most of the runtime advantage of frequency-domain block processing, a favorable trade-off that is unavailable to either conventional method alone.

The \emph{in vivo} runtimes (Table~\ref{tab:invivo_metrics}) compare COG RDA with RDA RTB and with DAS RTB, which was implemented and optimized by the scanner manufacturer.
RTB beamformers synthesize SA images directly from clinical focused-transmit and diverging-wave sequences without explicit matrix inversion, at a computational complexity of the same order as their multistatic counterparts \cite{jakovljevic2026power}.
The measured DAS RTB and RDA RTB runtimes are in agreement with those reported in \cite{jakovljevic2026power} for harmonic \emph{in vivo} imaging on the same CPU.

\subsection{Scalability to High-Element-Count Matrix Arrays for Volumetric Imaging}

COG decomposition of the multistatic reconstruction problem has the greatest practical impact for high-element-count 2-D arrays, such as PMUT and CMUT apertures \cite{moisello2024pmut,qiu2015piezoelectric}.
Standard multistatic RDA requires a 4-D FFT over the full transmit-receive tensor, with $\mathcal{O}(N^4)$ space complexity (Table~\ref{tab:complexity_summary}),
which makes it impractical for matrix arrays with more than $\sim$1000 elements, including the $32\times32$ array used in this work.
COG RDA requires only a 2-D FFT per offset gather (Section~\ref{subsec:cog_rda_2d}), reducing space complexity by a factor of $N^2$ while preserving standard RDA's $N^2$-fold reduction in computations relative to 3-D DAS (Table~\ref{tab:complexity_summary}).

On the $32\times32$ array, the $32\times$ speedup of GPU-based COG RDA over 3-D DAS translates into a $67\times$ energy saving, together with lower GPU power draw, utilization, and temperature increase (Fig.~\ref{fig:3d_gpu_resource_bars}).
The exception is peak GPU memory, which is approximately $4\times$ higher for COG RDA because the reconstructed volume and intermediate offset-gather buffers are held on the GPU.
Even so, the lower power draw and heat generation of COG RDA make it well suited to real-time volumetric beamforming on power- and thermally-constrained platforms, including battery-powered systems built around high-element-count PMUT/CMUT arrays.

\subsection{Extending COG Beamforming to Virtual-Source Acquisitions}

The multistatic datasets used to validate COG RDA are not routinely acquired in clinical practice, and were recovered here from focused-transmit acquisitions using REFoCUS \cite{bottenus2017recovery,ali2019extending}.
However, the matrix inversion in REFoCUS grows in time and especially memory cost with the number of transmits and receive channels, offsetting part of the memory savings that motivate COG beamforming.
A more direct route is to formulate COG beamforming natively in virtual source (VS) and receive element coordinates, building on our RTB formulation of RDA, which applies the multistatic signal model directly to VS/receive element pairs \cite{jakovljevic2026power}.
Grouping these pairs by common offset would extend COG RDA to clinically relevant focused-transmit and diverging-wave acquisitions without explicit matrix inversion and, for matrix arrays, would combine naturally with Explososcan sequences for real-time volumetric imaging.
Formulating and validating COG beamforming directly for VS acquisitions is a direction of ongoing work.

\section{Conclusion}
\label{sec:conclusion}
We proposed common offset gather (COG) beamforming, which reduces the memory requirements of frequency-domain multistatic beamforming by beamforming transmit-receive element pairs with a common offset independently, using a bistatic formulation.
Combined with the range-Doppler algorithm, COG RDA reduces space complexity by a factor of $N$ for 1-D arrays and $N^2$ for 2-D arrays relative to conventional multistatic RDA, while preserving the computational advantage of frequency-domain beamforming over DAS.
For linear arrays, COG RDA achieved image quality comparable to DAS and conventional RDA, while requiring two orders of magnitude less memory than conventional RDA and beamforming up to $25\times$ faster than DAS.
For a $32\times32$-element matrix array, where conventional RDA would require a 4-D FFT, GPU-based COG RDA reconstructed volumes $32\times$ faster and with $67\times$ less energy than 3-D DAS, with comparable point-target resolution and sidelobe levels.
These gains make COG RDA well suited to real-time volumetric imaging with high-element-count arrays on power- and thermally-constrained platforms.
In the future, we plan to formulate COG beamforming directly for virtual-source acquisitions, extending these advantages to focused-transmit and diverging-wave sequences without the need for REFoCUS.

\section*{Acknowledgment}
The authors would like to thank General Electric Healthcare for their in-kind and technical support.
Many thanks also go to Rajan Gurjar, Tanwin Chang, and Robert Haupt from MIT Lincoln Laboratory for insightful discussions about potential applications of this work to non-contact laser ultrasound.
Special thanks go to Shuangshuang (Selavie) Liu for inspiring conversations about this topic and about science in general.

\appendices

\section{POSP Derivation for 1-D Bistatic RDA}
\label{app:posp_1d}

For fixed offset $h$, the Fourier transform of $s_0(t,m,h)$ in \eqref{eq:signal_mo} over $m$ is a stationary-phase integral with phase
\begin{equation}
\Theta(m) = -j\tfrac{4\pi f_0}{c}\bigl(\sqrt{R_0^2+h^2} + \tfrac{R_0^2\,m^2}{2(R_0^2+h^2)^{3/2}}\bigr) + j2\pi m k_m.
\label{eq:phase_1d_app}
\end{equation}
Setting $d\Theta/dm = j\frac{4\pi f_0}{c}\frac{R_0^2}{(R_0^2+h^2)^{3/2}}\,m - j2\pi k_m = 0$ gives $m = k_m/K_m$, with the midpoint modulation rate
\begin{equation}
K_m = \frac{2f_0}{c}\,\frac{R_0^2}{\bigl(R_0^2+h^2\bigr)^{3/2}}.
\label{eq:Km_app}
\end{equation}
Substituting $m = k_m/K_m$ into \eqref{eq:phase_1d_app} yields the RD domain signal model in \eqref{eq:signal_rd}--\eqref{eq:travel_time_rd}, with phase
$\Theta(k_m) = -j\frac{4\pi f_0}{c}\sqrt{R_0^2+h^2} + j\pi k_m^2/K_m$.

\section{POSP Derivation for 2-D Bistatic RDA}
\label{app:posp_2d}

For a scatterer at depth $R_0$ and offset $\mathbf{h} = (h_x, h_y)^\top$, the exact travel path is
\begin{equation*}
f(\mathbf{m}) = R_{tx} + R_{rx} = \sqrt{R_0^2 + \|\mathbf{m}+\mathbf{h}\|^2} + \sqrt{R_0^2 + \|\mathbf{m}-\mathbf{h}\|^2}.
\end{equation*}
A second-order Taylor expansion around $\mathbf{m} = \mathbf{0}$ gives
\begin{equation}
f(\mathbf{m}) \approx 2\sqrt{D} + \mathbf{m}^\top\!\left(\frac{\mathbf{I}}{\sqrt{D}} - \frac{\mathbf{h}\mathbf{h}^\top}{D^{3/2}}\right)\!\mathbf{m}
  = 2\sqrt{D} + \mathbf{m}^\top \mathbf{A}\,\mathbf{m},
\label{eq:travel_path_2d_app}
\end{equation}
where $D = R_0^2 + \|\mathbf{h}\|^2$ and $\mathbf{A}$ is defined in \eqref{eq:A_matrix}.
The phase of the 2-D Fourier integral over $\mathbf{m}$ is
\begin{equation*}
\Theta(\mathbf{m}) = -j\frac{2\pi f_0}{c}\,f(\mathbf{m}) + j2\pi\,\mathbf{k}_m^\top\mathbf{m},
\end{equation*}
and setting $\nabla_\mathbf{m}\Theta = 0$ gives the stationary point
\begin{equation*}
\frac{2f_0}{c}\,\mathbf{A}\,\mathbf{m} = \mathbf{k}_m
\quad \Rightarrow \quad
\mathbf{m} = \frac{c}{2f_0}\,\mathbf{A}^{-1}\mathbf{k}_m.
\end{equation*}
By the Sherman-Morrison formula, using $D - \|\mathbf{h}\|^2 = R_0^2$,
\begin{equation}
\mathbf{A}^{-1} = \sqrt{D}\left(\mathbf{I} + \frac{\mathbf{h}\mathbf{h}^\top}{D - \|\mathbf{h}\|^2}\right)
  = \sqrt{D}\left(\mathbf{I} + \frac{\mathbf{h}\mathbf{h}^\top}{R_0^2}\right).
\label{eq:A_inv_app}
\end{equation}
Substituting the stationary point into $\Theta(\mathbf{m})$ gives
\begin{align}
\Theta(\mathbf{k}_m) &= -j\frac{2\pi f_0}{c}\left(2\sqrt{D} + \frac{c^2}{4f_0^2}\,\mathbf{k}_m^\top\mathbf{A}^{-1}\mathbf{k}_m\right) \nonumber \\
  &\quad + j\frac{\pi c}{f_0}\,\mathbf{k}_m^\top\mathbf{A}^{-1}\mathbf{k}_m \nonumber \\
&= -j\frac{4\pi f_0}{c}\sqrt{D} + j\frac{\pi c}{2f_0}\,\mathbf{k}_m^\top\mathbf{A}^{-1}\mathbf{k}_m,
\label{eq:phase_rd_2d}
\end{align}
which yields the RD domain signal model in \eqref{eq:signal_rd_2d}--\eqref{eq:travel_time_rd_2d}.

\bibliographystyle{IEEEtran}
\bibliography{abbrev,bibdata}

\end{document}